\documentclass[aps,prb,twocolumn,floats,showpacs]{revtex4}

\usepackage{epsfig}

\usepackage{amsmath}

\usepackage{amssymb}

\begin{document}

\title{Tunneling through triple quantum dots with mirror symmetry}
\author{$^1$T. Kuzmenko, $^2$K. Kikoin, and $^{2,3}$Y. Avishai}
\affiliation{$^1$The Rudolph Peierls Center for Theoretical
Physics, University of Oxford, 1 Keble Road, Oxford OX1 3NP,
United Kingdom \\
$^2$Department of Physics, Ben-Gurion University of
the Negev, Beer-Sheva 84105, Israel\\
$^3$ Ilse Katz Center for Nano-Technology,  Ben-Gurion University
of the Negev, Beer-Sheva 84105, Israel\\
}
\date{\today}

\begin{abstract}
Indirect exchange interaction between itinerant electrons and
nano-structures with non-trivial geometrical configurations
manifests a plethora of unexpected results. These configurations
can be realized either in quantum dots with several potential
valleys or in real complex molecules with strong correlations.
Here we demonstrate that the Kondo effect may be suppressed under
certain conditions in triple quantum dots with mirror symmetry at
odd electron occupation. First, we show that the indirect exchange
has ferromagnetic sign in the ground state of triple quantum dot
in a two-terminal cross geometry for electron occupation $N=3$.
Second, we show that for electron occupation $N=1$ in
three-terminal fork geometry the zero-bias anomaly in the tunnel
conductance is absent (despite the presence of Kondo screening)
due to special symmetry of the dot wave function.
\end{abstract}
\maketitle

\section{Introduction}

 Many-particle effects in quantum tunneling through quantum dots
are extensively discussed in the current literature (see, e.g.,
recent reviews \cite{KKAv,GPr}). Single-electron tunneling under
the conditions of strong Coulomb blockade is accompanied by
co-tunneling processes with spin-reversal, which involve dynamical
screening effects similar to the celebrated Kondo-scattering in
magnetically doped metals. The pertinent effect was predicted and
observed in single-well quantum dots with odd electron occupation,
where the electrons confined in the well are represented by the
non-compensated spin 1/2 of an electron on the highest occupied
discrete level. This generic pattern may be enriched in many ways,
in particular, by studying the tunneling through complex quantum
dots containing two or three valleys.

In this paper we are interested in specific physical properties of
electron tunneling through complex quantum dots containing several
potential valleys with essentially different capacitances.
Originally, the idea of coupling several nanoobjects having strong
and weak Coulomb interactions is formulated  in the context of
electron and spin structure of complex molecules (e.g.,
lantanocenes, containing strongly correlated f-electrons
hybridized with weakly correlated molecular orbitals occupied by
p-electrons). It is noticed \cite{Neuf91} that the energy
difference between the singlet ($S$) ground state and triplet
($T$) excited  state of a molecule with even number of electrons
$N=N_f+N_p$ is anomalously small, so that the triplet excitation
affects the magnetic response of the system. The simplest
artificial analog of this system is an asymmetric double quantum
dot (composed of big and small dots) with even occupation (e.g.,
$N=2$) and charging energies $Q_s \gg Q_b$ for small and big dots.
When this double quantum dot is coupled with the metallic leads
via the big dot (the T-shaped geometry) then lead-dot electron
tunneling may induce an $S\to T$ crossover \cite{KAv}: the energy
$E_T$ of the triplet state becomes lower than the energy $E_S$ of
the singlet state, and the Kondo-regime may show up. In case of
odd occupation $N=1$, the Kondo-Fano regime is relevant, where the
Kondo effect induced by an electron localized in the small side
dot affects the electron tunneling through the weakly correlated
big dot \cite{Fano}. If the leads are connected via the small dot,
the big dot plays the role of additional reservoir for Kondo
screening in case of odd $N$ and the two-channel Kondo effect may
be realized under certain conditions.\cite{OG,PBLD}

The more complicated model of asymmetric triple quantum dot (TQD)
with a small dot sandwiched between two big dots was considered in
Ref. \cite{KuKA} for odd and even occupations $N=3,4$. It was
shown that the Kondo regime is accessible in both cases. Moreover,
the TQD with even occupation demonstrates  $SO(n)$ symmetry.

\begin{figure}[htb]
\centering
\includegraphics[width=70mm,height=40mm,angle=0]{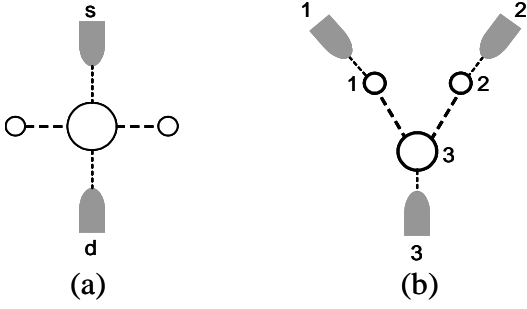}
\caption{Triple quantum dots in "cross" (a) and "fork" (b)
geometry.}\label{TQD}
\end{figure}

Recently, a mirror symmetric TQD (see figure \ref{TQD}) in an
"open" regime (near the Coulomb blockade peak) were studied both
experimentally \cite{Marcus} and theoretically \cite{Vava}. In
this case the big "mesoscopic" dot is connected with two small
dots and with metallic reservoirs. As a result, an indirect RKKY
exchange interaction between the spins in the couple of small dots
occurs, and its sign may be controlled by changing the parameters
of the large central dot by applying an external magnetic field.

Here we focus on a mirror symmetric TQD in the "closed" regime,
that is, the valley between Coulomb blockade peaks where the total
number of electrons in the TQD is fixed. Our main results are: (i)
in the cross geometry (Fig. \ref {TQD}a) with electron occupation
$N=3$, such a dot possesses an unusual property: {\it the indirect
exchange tunneling constant between the big dot and the leads is
ferromagnetic}; the Kondo tunneling is therefore absent although
the TQD behaves as a local moment; (ii) in a "fork" configuration
(Fig. \ref{TQD}b) with electron occupation $N=1$, TQD exhibits two
different tunneling regimes: depending on the gate voltages,  the
Kondo regime may be observed as a zero bias anomaly or as a finite
bias anomaly in the tunnel conductance.

\section{Triple quantum dot in a cross geometry}

The TQD in the cross geometry (figure \ref {TQD}a) is
composed of left $l$, center $c$, and right $r$ dots, with
corresponding levels and charging energies $\epsilon_{j},Q_{j},
{j=l,c,r}.$ It is modeled by the Anderson Hamiltonian
\begin{equation}
H=H_{d}+H_{lead}+H_t, \label{H-sys}
\end{equation}
containing the terms describing the dot, two leads and the
dot-lead tunneling, respectively. The first term is
\begin{eqnarray}
H_{d}&=&\sum_{j=l,c,r}\sum_{\sigma}\epsilon_{j}d^\dagger_{j
\sigma}d_{j\sigma
}+\sum_{j}Q_jn_{j\uparrow}n_{j\downarrow}\nonumber\\
&+&W\sum_{j=l,r}\sum_{\sigma}(d^\dagger_{j \sigma}d_{c\sigma
}+H.c.).\label{H-dot}
\end{eqnarray}
The parameters $Q_j$ are chosen in such a way that for electron
occupation $N=3$, and in the absence of inter-dot tunneling
($W=0$), each dot is occupied by one electron. In the
mirror-symmetric case such configuration is realizable provided
\begin{eqnarray}
&& \epsilon_l=\epsilon_r\equiv\epsilon_s,~~~
\Delta=\epsilon_c-\epsilon_s>0, \nonumber \\
&& Q_l=Q_r\equiv Q_s \gg Q_c ~.
\end{eqnarray}
At finite $W$ charge transfer from the central dot to the side
dots is possible, but the double occupation of the side valleys is
still suppressed by strong Coulomb blockade $Q_s\gg W$. In the
charge sector $N=3$ and for the mirror symmetric configuration the
lowest energy states are two spin doublets, (even and odd relative
to the $l\leftrightarrow r$ permutation), a spin quartet state,
and a doubly degenerate charge transfer exciton (with two
electrons in the central dot). The corresponding energies are
\begin{eqnarray}
E_{Du}&=&2\epsilon_s+\epsilon_c-3W^2/\Delta,\nonumber \\
E_{Dg}&=&2\epsilon_s+\epsilon_c-W^2/\Delta,\nonumber \\
E_{Q}&=&2\epsilon_s+\epsilon_c,  \label{spin-mul}\\
E_{Ex}&=&\epsilon_s+2\epsilon_c+2W^2/\Delta .\nonumber
\end{eqnarray}
Here the inequality $W/\Delta\ll 1$ is assumed to be valid. The
eigenfunctions of the doublet and quartet states (which
predetermine the structure of the effective exchange Hamiltonian,
see below) are:
\begin{eqnarray}
 |D_u\sigma\rangle &=&
\bigg[{\cos\theta_u}
      \frac{(b^{\dag}_{cr}d^{\dag}_{l\sigma}-
       b^{\dag}_{cl}d^{\dag}_{r\sigma})}{\sqrt{3}}+\nonumber
 \\&+&
      \sin\theta_u b^{\dag}_{cc}
      (d^{\dag}_{l\sigma}-d^{\dag}_{r\sigma})
 \bigg]|0\rangle,
 \nonumber \\
 |D_g\sigma\rangle &=&
 \bigg[
      {\cos\theta_g}
      b^{\dag}_{lr}d^{\dag}_{c\sigma}-
      \sin\theta_gb^{\dag}_{cc}
      (d^{\dag}_{l\sigma}+d^{\dag}_{r\sigma})
 \bigg]|0\rangle, \nonumber\\
\left| Q, {+\frac{3}{2}}\right\rangle
&=&d^{+}_{l\uparrow}d^{+}_{c\uparrow}d^{+}_{r\uparrow}\vert
0\rangle , \;\;\ \left | Q, -\frac{3}{2}\right\rangle
=d^{+}_{l\downarrow}d^{+}_{c\downarrow}d^{+}_{r\downarrow}\vert
0\rangle , \nonumber \\
\left |Q, {+\frac{1}{2}}\right\rangle
&=&\frac{1}{\sqrt{3}}\sum_{\langle ijk\rangle}
d^{+}_{i\uparrow}d^{+}_{j\uparrow}d^{+}_{k\downarrow}\vert
0\rangle , \nonumber\\
\left |Q, {-\frac{1}{2}}\right\rangle
&=&\frac{1}{\sqrt{3}}\sum_{\langle ijk\rangle}
d^{+}_{i\downarrow}d^{+}_{j\downarrow}d^{+}_{k\uparrow}\vert
0\rangle . \label{func}
\end{eqnarray}
Here $b^{\dag}_{ij}=[d^{\dag}_{i\uparrow}d^{\dag}_{j\downarrow}-
d^{\dag}_{i\downarrow}d^{\dag}_{j\uparrow}(1-\delta_{ij})]/{\sqrt
2}$, and rotation angles $\theta_u=\arcsin(\sqrt{3}W/\Delta),$
$\theta_g=\arcsin(W/\Delta).$

 The two other terms in Eq.
(\ref{H-sys}) are the band Hamiltonian describing the electrons in
the leads
\begin{eqnarray}
H_{lead}&=&\sum_{a=s,d}\sum_{k\sigma}\epsilon_{ak}c^\dagger_{ak
\sigma}c_{ak\sigma}, \label{H-l}
\end{eqnarray}
and the tunneling Hamiltonian
\begin{eqnarray}
H_t=\sum_{ak\sigma}(V_a c^\dagger_{ak\sigma} d_{c\sigma}+
H.c.).\label{H-tun}
\end{eqnarray}

Following the standard procedure, one derives an indirect exchange
interaction between lead and dot electrons by means of the
Schrieffer-Wolff (SW) transformation. For a composite quantum dot,
this is accomplished in terms of spin eigenstates
 of $H_d$ defined in Eqs. (\ref{spin-mul}),(\ref{func}).

 Thus in the SW regime the main contribution to the lead-dot tunneling
is given by the components $\sim \cos\theta_{g,u}$. The quartet
state has the usual structure prescribed by a standard Young
tableau for three electrons, ($\langle ijk\rangle$ indicates the
cyclic permutation of three sites $lcr$).

After performing the SW-like canonical transformation, one gets
the effective exchange Hamiltonian
\begin{equation}\label{Hu}
H_{ex}= J_{u}{\bf S}_u\cdot {\bf s},
\end{equation}
where ${\bf S}_u$ is the spin $1/2$ vector operator with
components $S^+_u=|u\uparrow\rangle\langle u\downarrow |$,
$S^z_u=(|u\uparrow\rangle\langle u\uparrow |-
|u\downarrow\rangle\langle u\downarrow|)/2$, and ${\bf s}$ is the
spin operator of lead electrons defined as ${\bf s}=\sum_k
c^\dag_{ek\sigma}\hat \tau c_{ek\sigma'}$; $\hat \tau$ is the set
of Pauli matrices, $c_{e\sigma}$ is the even combination of lead
electron annihilation operators (the odd one is excluded from the
tunneling Hamiltonian by standard rotation\cite{GR}). Remarkably,
the exchange constant $J_u$ has {\it ferromagnetic} sign,
\begin{equation}\label{Ju}
J_u=-\frac{2\cos^2{\theta_u}V^2}{3}
\left(\frac{1}{|\epsilon_c|}+\frac{1}{|\epsilon_c+Q_c|}\right)
\end{equation}
($|\epsilon_c|$ is the position of the central dot level relative
to the Fermi energy of the leads, source and drain contacts are
assumed to be equivalent, $V_s=V_d=V$). The reason for an
unconventional sign of the exchange interaction is that only one
of three electrons in the TQD is involved in exchange interaction
with the leads, and the overlap of the two other electrons  wave
function entering the state $|D_{u}\sigma\rangle$ gives the factor
$- 1$. Thus in this geometry we encounter a unique situation,
where the Kondo screening is ineffective for a quantum dot with
{\it odd occupation}.

\begin{figure}[htb]
\centering
\includegraphics[width=70mm,height=60mm,angle=0]{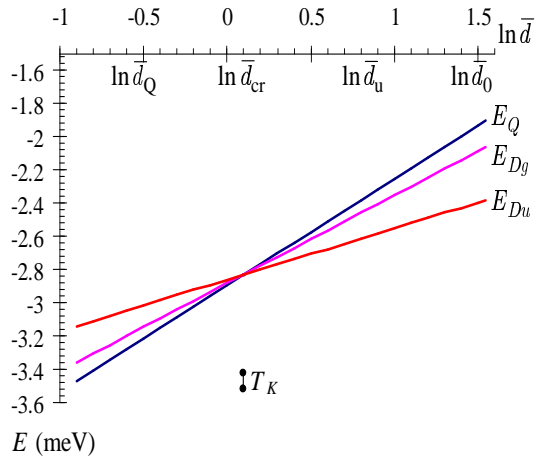}
\caption{Haldane flow diagram for the levels $E_{\Lambda}$of
(\ref{scinv}), $\bar d=\pi \bar D/{\Gamma_Q}$. Energy is measured
in meV units.} \label{hald}
\end{figure}

Yet, a crossover from ferromagnetic to anti-ferromagnetic scenario
is feasible. Indeed, the level spacing in the spin multiplet
(\ref{spin-mul}) is governed by the parameter $W/\Delta$. If this
spacing is small enough, the renormalization of these levels due
to lead-dot co-tunneling becomes relevant in the RG flow equations
along with the exchange screening in the framework of Haldane
renormalization procedure \cite{Hald}. Accordingly, the energy
levels (\ref{spin-mul}) are renormalized as a result of
integrating out the band edges and shrinking the band width from
its bare value $D_0$ to a smaller value $D$ comparable with
$|\epsilon_c|$). The corresponding RG invariant is,
\begin{equation}\label{scinv}
E_\Lambda^*=E_\Lambda(D)-\pi^{-1}\Gamma_\Lambda\ln (\pi
D/\Gamma_\Lambda)
\end{equation}
with tunneling rates $\Gamma_{D(u,g)}\approx \pi\rho_0\cos^2
\theta_{(u,g)}V^2$ and $\Gamma_{Q}\approx \pi\rho_0 V^2$. Due to
the hierarchy  $\Gamma_Q> \Gamma_{Dg}>\Gamma_{Du}$, a level
crossing is feasible (see Fig. \ref{hald}).
Physically, it implies a crossover from the non-Kondo
(ferromagnetic exchange) regime to the under-screened Kondo regime
with a pronounced maximum of the conductance around the degeneracy
point. The  parameters $W$ and $\Delta$, which determine the
initial conditions (\ref{spin-mul}) for the flow equations
(\ref{scinv}) can be controlled by gate voltages. Varying these
initial conditions, one may tune the region of crossover to the SW
regime at the point
\begin{equation}\label{schwo}
\bar D \approx E_\Lambda(\bar D).
\end{equation}
For $D<\bar D$ the properties of the system are determined by the
SW Hamiltonian. The effective Hamiltonian (\ref{Hu}) may be
written for $\bar D = \bar D_u$ (marked on the horizontal axis of
Fig. \ref{hald}). If the condition (\ref{schwo}) is fulfilled in
the vicinity of the crossing point ($\bar D = \bar D_{cr}$), the
exchange Hamiltonian (instead of (\ref{Hu})) acquires the form
\begin{equation}\label{swal}
H_{SW}=J_u {\bf S}_u\cdot {\bf s}+J_g {\bf S}_g\cdot {\bf s}+J_Q
{\bf S}_Q\cdot {\bf s}+ J_R {\bf R}\cdot {\bf s},
\end{equation}
expressed in terms of operators for localized spin 1/2, ${\bf
S}_{u,g}$, spin 3/2, ${\bf S}_{Q}$, and the vector ${\bf R}$ which
induces transitions between the quartet $|Q\rangle$ and
the doublet $|D_u\rangle$. There are no transitions between
$|Q\rangle$ and $|D_g\rangle$ since these states have different
$l-r$ parity.

\begin{figure}[htb]
\centering
\includegraphics[width=60mm,height=45mm,angle=0]{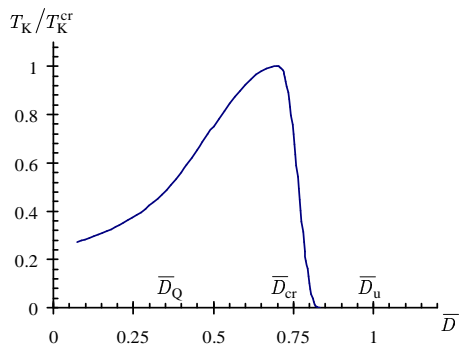}
\caption{Evolution of Kondo temperature as determined by scaling
equations (\ref{sc-eqs}).} \label{kondot}
\end{figure}

At the point $\bar D_{cr}$ the degeneracy of spin state of the TQD
is maximal, corresponding to the symmetry $SU(2)\times SU(2)
\times SU(2) $. Both to the right and to the left of this point
some of the states in the spin multiplets are quenched at $T\to
T_K$ and $T_{K}$ depends on the energy gaps $\Delta_{Qg}=E_Q(\bar
D)-E_{Dg}(\bar D)$ and $\Delta_{gu}=E_{Dg}(\bar D)-E_{Du}(\bar
D)$. To find the function $T_K(\Delta_{Qg},\Delta_{gu})$, one
should solve the scaling equations for the coupling constants in
the Hamiltonian $H_{SW}$,
\begin{eqnarray}
&&\frac{dj_u}{d\ln d}=-[j_u^2+2j_R^2],\ \ \ \frac{dj_g}{d\ln
d}=-j_g^2,\label{sc-eqs}\\
&& \frac{dj_Q}{d\ln d}=-[j_Q^2-j_R^2], \ \ \  \frac{dj_R}{d\ln
d}=-\frac{j_R}{4}(5j_Q-j_u),\nonumber
\end{eqnarray}
where $j_a=\rho_0J_a$ $(a=g,u,Q,R)$, and $\rho_0$ is the density
of states which is assumed to be constant. The procedure is
self-consistent because $T_K$ itself predetermines the
characteristic energy interval for states in the spin Hamiltonian
involved in its formation.
Varying $\bar D$ in Fig. \ref{hald}, from the ferromagnetic
non-Kondo regime ($\bar D \sim \bar D_u$) to the crossing point
$\bar D_{cr}$, one reaches the point $T_K >0$ which arises due to
influence of the excited states $E_Q(\bar D)$ and $E_{Dg}(\bar
D)$. Just then, $T_K$ sharply increases, reaching its maximum
value in the point of maximum degeneracy $\bar D_{cr}$. Moving
further to the left, the level $E_{Du}$ freezes out. This means
that the vector ${\bf R}$ in the Hamiltonian (\ref{swal}) does not
contribute anymore to Kondo co-tunneling, and the Kondo effect is
determined by the pair of states $E_{Q}$ and $E_{Dg}$, with the
dynamical symmetry of TQD being $SU(2)\times SU(2)$. Further
decrease of $\bar D$ eventually results in the quenching of
$E_{Dg}$.  The system then exhibits an under-screened Kondo regime
of a localized spin 3/2 moment. Fig. \ref{kondot} illustrates
these crossover effects on $T_{K}$.

\begin{figure}[htb]
\centering
\includegraphics[width=70mm,height=35mm,angle=0]{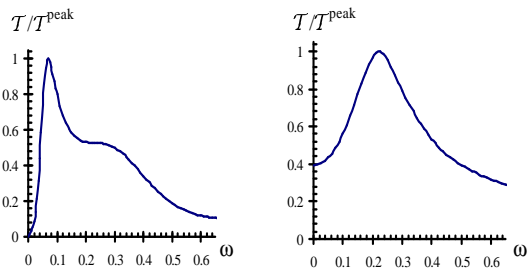}
\caption{Tunnel transparency of TQD in a doublet (left panel) and
quartet (right panel) ground states. ${\cal T}_{peak}$ is the
maximum value of ${\cal T}(\omega)$.} \label{transp}
\end{figure}

The evolution of $T_K$ is reflected in the behavior of tunnel
transparency and conductance as a function of energy. Far enough
to the left and to the right of the crossing point, this behavior
is stepwise. When the dot is in the doublet ground state ($\bar D
\sim \bar D_u$), the spin multiplet as a whole contributes to the
transparency at high energies $\omega \sim
\Delta_{Qg}+\Delta_{gu}$. With decreasing $\omega$ the quartet
$E_Q$ and the even doublet $E_{Dg}$ energies freeze out in this
order (they are not renormalized anymore). Eventually, Kondo
tunneling is quenched at low energies, so that the zero bias
anomaly (ZBA) has a shape of a dip. On the other hand, in the
crossover regime, the ZBA follows the conventional Kondo peak.
Finally, the structure of the peak at the regime ${\bar D} \sim
{\bar D}_Q$ is even more complicated. Within the framework of our
approach we may describe the evolution of transparency for $T
> T_K$ where it is approximately described by the simple relation
${\cal T}(\omega)\sim \ln^{-2}[T/T_K(\omega)]$. The resulting
curve is shown in Fig. \ref{transp}.

This type of non-universal behavior of $T_K$ is known in the
theory of strongly correlated quantum dots with {\it even}
occupation, where the singlet-triplet level crossing usually
occurs \cite{KAv,KuKA,Hof04}. The novelty of the present scenario
is that it is manifested in quantum dot with {\it odd} occupation
where the absence of Kondo effect occurs due to ferromagnetic
exchange coupling with the localized spin doublet. The
non-universality of $T_{K}$ occurs as this ferromagnetic exchange
competes with two anti-ferromagnetic exchange interactions (with
doublet and quartet localized moments), so that, in some sense,
one deals with a "three-stage" Kondo effect. Thus, we have
completed our discussion pertaining to the cross-shaped TQD of
figure \ref{TQD}a.

\section{Triple quantum dot in a fork geometry}

 It is then natural to expect peculiar features of Kondo
tunneling also in a mirror symmetric TQD in the fork geometry
(Fig. \ref{TQD}b) with $Q_{l,r} \gg Q_{c}$. The dots and leads are
labeled 1,2,3 and each dot is attached to its own lead. In case of
$N=3$,   the exchange coupling $J_{3}$ between the central dot and
its adjacent lead is ferromagnetic in accordance with Eq.
(\ref{Ju}), whereas those for the two other (small) dots
($J_{1},J_{2}$) are antiferromagnetic. Besides, there are also
non-diagonal exchange couplings $J_{ij}=J_{ji}$. All these are
coupled within a system of RG flow equations. A question now
arises, whether it is possible to find a regime where the Kondo
resonance arises only in dots 1,2, whereas dot 3 remains Kondo
inactive?

To answer this question, we calculate $T_K$ and $G$ within the
same scheme as in the preceding section for a system described by
the Hamiltonian (\ref{H-sys}) with the tunneling term
\begin{eqnarray}
H_t=\sum_{j=1}^{3}\sum_{k\sigma}(V_j c^\dagger_{jk\sigma} d_{j\sigma}+
H.c.)\label{H-tunf}
\end{eqnarray}
instead of (\ref{H-tun}). The effective exchange Hamiltonian of
this system is obtained in the same way as (\ref{swal}). The
mirror $l-r$ symmetry entails $V_1=V_2\neq V_3$. If the
ground-state consists of the odd-parity doublet $E_{D_u}$, the
corresponding SW Hamiltonian has the form
\begin{eqnarray}
H_{SW}&=&\sum_{i=1}^3J_i{\bf S}\cdot {\bf s}_i+J_{12}{\bf S}\cdot ({\bf
s}_{12}+{\bf s}_{21})\label{Hspin}\\
&+&J_{13}{\bf S}\cdot ({\bf s}_{13}+{\bf s}_{31})+ J_{23}{\bf
S}\cdot ({\bf s}_{23}+{\bf s}_{32}).\nonumber
\end{eqnarray}
Here the exchange constant $J_3<0$ is the same as $J_u$
(\ref{Ju}), whereas $J_{1,2}>0$. For $\cos \theta_u \approx 1$,
these constants are,
\begin{eqnarray}\label{coupl-Ja}
 J_1 &=&J_2=\frac{4}{3}\frac{V_1^2}{|\epsilon_s|}, \nonumber\\
  J_3 &=&-\frac{2}{3}\left(\frac{V_3^2}{|\epsilon_c|}+
  \frac{V_3^2}{|\epsilon_c+Q_c|}\right),\label{J}\\
J_{12}
&=&-\frac{3}{2}\frac{(V_1W)^2}{(\epsilon_c+Q_c-\epsilon_s)^2|\epsilon_s|}~, \nonumber \\
J_{13} & = & J_{23} =-\frac{1}{4}\frac{W
V_1V_3}{\epsilon_c+Q_c-\epsilon_s} \left(\frac{1}{|\epsilon_s|}+
\frac{1}{|\epsilon_c|}\right). \nonumber
\end{eqnarray}
The system of RG flow equations for the Hamiltonian (\ref{Hspin})
now reads,
\begin{eqnarray}
 \frac{dj_1}{d\ln d} &=&
 -\left[j_1^2+j_{13}^2+j_{12}^2\right],
\nonumber \\
  \frac{dj_3}{d\ln d} &=&
 -\left[j_3^2+2j_{13}^2\right],
 \label{scal} \\
 \frac{dj_{12}}{d\ln d} &=&
 -2j_1j_{12},
 \nonumber\\
  \frac{dj_{13}}{d\ln d} &=&
 -j_{13}(j_1+j_3). \nonumber
 \end{eqnarray}

A somewhat unexpected results following the analysis of these
equations: The coupling constants related to channel 3 become
relevant in spite of their negative initial values (\ref{J}). The
Kondo temperature is
\begin{eqnarray}\label{T1}
T_K&=&{\bar D}\exp{\left(-\frac{2}{j_+
+\sqrt{j_-^2+6j_{13}^2}}\right)},
\end{eqnarray}
where $j_+=j_1+j_{12}+j_{3}$, $j_-=j_1+j_{12}-j_{3}$, so that the
Kondo resonance arises in all non-diagonal channels and the TQD
looses its "exotic" properties. The Kondo transparency ${\cal
T}_{ij}(\omega)$ may be calculated for any pair of electrodes
$(ij)$. It is a step-wise function in accordance with multistage
Kondo screening process, but  no anomalous "freezing out" of Kondo
effect similar to that shown in the left panel of Fig.
\ref{transp} is expected in this case.

   It is appealing, however, to exploit other specific properties
of the TQD in the fork configuration.  The remarkable feature of
the mirror asymmetric TQD is that it can be viewed as a quantum
pendulum \cite{Sarag,Lehur}. This means that the superposition of
two degenerate states (13) and (23) can be considered as a sort of
mesoscopic resonating valence bond (RVB). In these two papers
various manifestations of spin entanglement at even occupation
$N=4,2$ were investigated. Below we will discuss how this property
is manifested in the Kondo regime in the odd occupation charge
sector $N=1$.

So let us assume that the parameters (gate voltages) are tuned so
that the TQD is found in a Coulomb blockade valley corresponding
to the occupation sector $N=1$. Within the same approximation as
(\ref{spin-mul}) the lowest eigenstates $|\Lambda\rangle$ of the
Hamiltonian $H_{d}$ (\ref{H-dot}) for $N=1$ are the set of spin
doublets,
\begin{eqnarray}
|Db\sigma\rangle &=&
\bigg[\sin\theta~ d^{\dag}_{3\sigma} +
\cos\theta~\frac{d^{\dag}_{1\sigma}+d^{\dag}_{2\sigma}}{\sqrt{2}}
\bigg]|0\rangle,
 \nonumber \\
 |Dn\sigma\rangle & =& \bigg[\frac{d^{\dag}_{1\sigma}-d^{\dag}_{2\sigma}}
{\sqrt{2}} \bigg]|0\rangle,
 \label{func2} \\
  |Da\sigma\rangle & =& \bigg[-\cos\theta~ d^{\dag}_{3\sigma} +
\sin\theta~\frac{d^{\dag}_{1\sigma}+d^{\dag}_{2\sigma}}
{\sqrt{2}} \bigg]|0\rangle~,
 \nonumber
\end{eqnarray}
with $\sin \theta =\sqrt{2}W/\Delta \ll 1$. The corresponding
eigenfunctions are
\begin{eqnarray}
E_{Db}&=& \epsilon_s -2W^2/\Delta~,\nonumber \\
E_{Dn} &=& \epsilon_s~, \label{sps} \\
E_{Da}&=&\epsilon_c +2W^2/\Delta .\nonumber
\end{eqnarray}
Hence, the ground state is a bonding spin doublet $E_{Db}$, and
the eigenstates $|\Lambda\rangle$ may be interpreted as two even
and one odd RVB modes of a "spin pendulum".

In the low-energy subspace $\omega \ll W^2/\Delta$ the effective
spin Hamiltonian which describes the Kondo co-tunneling has the
same form as (\ref{Hspin}) derived above for $N=3$. Here, however,
all coupling constants are positive,
\begin{eqnarray}
&& J_{1}=J_{2}=J_{12}=\frac{V_1^2\cos^2 \theta}{2|\epsilon_s|}, ~~
J_{3}=\frac{V_3^2\sin^2 \theta}{|\epsilon_c|},\nonumber\\
&& J_{13}=J_{23}=\frac{V_1V_3\sin
\theta\cos\theta}{2\sqrt{2}}\left
(\frac{1}{|\epsilon_c|}+\frac{1}{|\epsilon_s|}\right), \label{J_1}
\end{eqnarray}
and the Kondo temperature is given by Eq. (\ref{T1}).

One may say that in a coherent Kondo-tunneling regime the system
demonstrates perfect entanglement: an electron entering dot 3 from
 lead 3, splits into two components in accordance with the
structure of the state $|Db\sigma\rangle$ and this entangled state
predetermines the total current $I_1 + I_2$ through the TQD in the
fork geometry. Of course, this statement is valid only at zero
temperature, and one may expect that thermal fluctuations are
detrimental for a coherent transport, but this effect is
suppressed as $(T/T_K)^2$ at low $T$.

\begin{figure}[htb]
\centering
\includegraphics[width=60mm,height=40mm,angle=0]{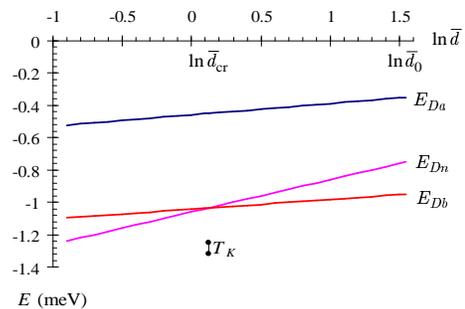}
\caption{Haldane flow diagram for the levels $E_{\Lambda}$ of TQD
in fork geometry, $\bar d=\pi \bar D/{\Gamma_n}$. Energy is
measured in meV units.} \label{hald2}
\end{figure}

The situation becomes even richer, due to the occurrence of soft
mode excitations (\ref{sps}): the odd state $|Dn\sigma\rangle$ may
be intermixed with the even state $|Db\sigma\rangle$ due to
co-tunneling process. This intermixing becomes relevant provided
$T_K \lesssim E_{Dn}-E_{Da}$. This inequality is, of course
invalid for the bare eigenstates (\ref{sps}), but the Haldane
renormalization of the spectrum similar to that described by Eq.
(\ref{scinv}) may result in softening of this mode. It might even
lead to level crossing provided the tunneling rate
$\Gamma_{n}=\rho_1V_1^2$ for the non-bonding state
$|Dn\sigma\rangle$ is higher than the rate
$\Gamma_{b}=\rho_1V_1^2\cos^2\theta+ \rho_3V_3^2\sin^2\theta$ for
the bonding state $|Db\sigma\rangle$. If the densities of states
are the same in all leads, $\rho_1=\rho_3$, then the condition
$\Gamma_{n}> \Gamma_{b}$ means $V_1>V_3$.  The RG flow
trajectories for this case are presented in Fig. \ref{hald2}.
Like in the cross geometry, the TQD acquires an additional
degeneracy in the critical region $\bar d \sim \bar d_{cr}.$
However, in this case the sources of degeneracy are the RVB
degrees of freedom. As a result, one encounters the problem of
Kondo effect due to an interplay between spin ${\bf S}$ and
pseudospin $\boldsymbol {\cal T}$, where the latter describes the
pendulum degrees of freedom. This problem was discussed in the
context of double quantum dots,\cite{Bord03} triple quantum dots
\cite{Saka05,KKA05} and molecular trimers chemisorbed on metallic
surfaces \cite{Affleck}. In this case the actual symmetry of the
TQD is $SU(4)$. To show this, we derive below the effective spin
Hamiltonian following the method offered in Ref.
\onlinecite{KKA05}.

It is useful to generalize the notion of localized spin operator
${ S}^i= |\sigma\rangle \hat \tau_i \langle\sigma'|$ (employing
Pauli matrices $\hat\tau_i~ (i=x,y,z)$) to
${S}^i_{\Lambda\Lambda'}= |\sigma\Lambda\rangle \hat \tau_i
\langle\sigma'\Lambda'|$, in terms of the eigenvectors
$|Db\sigma\rangle,|Dn\sigma\rangle$ from (\ref{func2}). Similar
generalization applies for the spin operators of the lead
electrons: ${ s}^i_{\lambda\lambda'}=\sum_{{\bf k}{\bf
k'}}c^\dag_{\lambda,{\bf{k}}\sigma}\hat \tau_i c_{\lambda',{\bf
k'}\sigma'}$. Here the index $\lambda$ denotes conduction
electrons in lead 3 ($\lambda=3$), and in leads 1,2 with
$\lambda=e,o$ corresponding to even and odd combinations of
conduction electron states in these leads,
\begin{equation}
c^\dag_{(e,o){\bf k}\sigma}=\frac{1}{\sqrt{2}}(c^\dag_{1,{\bf
k}\sigma} \pm c^\dag_{2,{\bf k}\sigma}).
\end{equation}
The pseudospin operators ${\boldsymbol {\cal T}}$ describing the
RVB mode are introduced as follows:
\begin{eqnarray}
{\cal T}^{+}&=&\sum_\sigma |Db\sigma\rangle\langle
Dn\sigma|, \ \ {\cal T}^{-}=[{\cal T}^{+}]^{\dagger},\label{taud}\\
{\cal T}^z&=&\frac{1}{2}\sum_\sigma\left(
|Db\sigma\rangle\langle Db\sigma|-|Dn\sigma\rangle\langle
Dn\sigma| \right).\nonumber
\end{eqnarray}
The five vector operators ${\bf S}_{\Lambda\Lambda'}$ and
${\boldsymbol {\cal T}}$ constitute the 15 generators of the
$SU(4)$ group.

Similarly, one may construct the pseudospin operators  for the
electrons in the leads (1,2):
\begin{eqnarray}
\tau^{+}&=&\sum_{k\sigma}
         c^{\dag}_{ek\sigma}c_{ok\sigma},~~~
\tau^{-}=[\tau^{+}]^{\dagger},\label{taub}\\
\tau_z &=& \frac{1}{2}
  \sum_{k\sigma}
  (c^{\dag}_{ek\sigma}
       c_{ek\sigma}-c^{\dag}_{ok\sigma}c_{ok\sigma}).\nonumber
\end{eqnarray}
The exchange Hamiltonian for TQD with spin and RVB degrees of
freedom is
\begin{eqnarray}\label{swfork}
H_{SW}=\sum_{\kappa\lambda\mu\rho}J_{\kappa\lambda\mu\rho}{\bf
S}_{\kappa\lambda}\cdot {\bf s}_{\mu\rho} + J_p \boldsymbol {\cal
T}\cdot{\boldsymbol \tau}
\end{eqnarray}
with $\kappa,\lambda=b,n$; $\mu,\rho=e,o$, and the coupling
constants
$J_{\kappa\lambda\mu\rho}=J_{\lambda\kappa\mu\rho}=J_{\kappa\lambda\rho\mu}$
are positive like in (\ref{J_1})
\begin{eqnarray}
 &&J_{bbee}=2J_{11},
   \ \ \ \
J_{bb33}=J_{33},   \nonumber
\\ && J_{bnoe}=\frac{V_1^2\cos{\theta}}{|\epsilon_s|} ,\ \ \
J_{nnoo}=\frac{V_1^2}{|\epsilon_s|},\label{Jnbeo}\\
&& J_{bb3e}=J_{13},\ \
 J_p=2J_{bbee}.\nonumber
\end{eqnarray}
The system of scaling equations has the  form:
\begin{eqnarray}
 \frac{dj_{b}}{d\ln d} &=&
 -\left[j_{b}^2+\frac{j_{bn}^2}{2}+j_{bn}j_p+{j_{13}^2}\right],
\nonumber \\
 \frac{dj_{n}}{d\ln d} &=&
 -\left[j_{n}^2+\frac{j_{bn}^2}{2}+j_{bn}j_p\right],
 \nonumber \\
 \frac{dj_3}{d\ln d} &=&
 -\left[j_3^2+{j_{13}^2}\right],
 \label{scal-f} \\
 \frac{dj_{13}}{d\ln d} &=&
 -j_{13}\left(j_{b}+j_3\right),
 \nonumber\\
 \frac{dj_{bn}}{d\ln d} &=&
 -\frac{1}{2}\left(j_{bn}+j_p\right)\left(j_{b}+j_{n}\right),
 \nonumber \\
 \frac{dj_p}{d\ln d} &=&
 -j_p^2, \nonumber
 \end{eqnarray}
where $j_b=j_{bbee}$, $j_n=j_{nnoo}$, $j_3=j_{bb33}$, $j_{13}=j_{bb3e}$
and $j_{bn}=j_{bnoe}$. From these equations we derive  the Kondo
temperature
\begin{equation}\label{kondof}
  T_K={\bar D}
  \exp\left\{
           {-\frac{2}
                  {j_{+}+
                  \sqrt{6j_{13}^2+(j_{bn}+j_p)^2+j_{-}^2}}}
      \right\},
\end{equation}
with $j_{+}=j_{b}+j_{n}+j_3$, $j_{-}=j_{b}-j_{n}-j_3.$ Like in the
cross geometry, one may manipulate $\bar D$ by changing the gate
voltages and scan the dependence $T_K(\bar D)$ similarly to Fig.
\ref{kondot}. This curve has a maximum in the critical point $\bar
D_{cr}$,  and the orbital degrees of freedom are frozen out in the
asymptotic regimes $\bar D \gg \bar D_{cr}$ and $\bar D \ll \bar
D_{cr}$. We deal in this case with a symmetry crossover $SU(2)\to
SU(4)\to SU(2)$. However, there is an important difference between
the two asymptotic $SU(2)$ symmetries.
\begin{figure}[htb]
\centering
\includegraphics[width=50mm,height=40mm,angle=0]{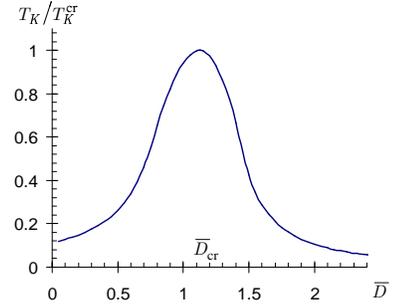}
\caption{Evolution of Kondo temperature, $T_K^{cr}$ as determined
by Eq.(\ref{kondof}).} \label{tk-fork}
\end{figure}

In the limit $\bar d \gg \bar d_{cr}$ the ground state is
$E_{db}$, all three dots are partially occupied in accordance with
the structure of the corresponding wave function
$|Db\sigma\rangle$ ( \ref{func2}). The terms $\sim J_{bbee}$,
$J_{bb33}$, $J_{bb3e}$  survive in the SW Hamiltonian
(\ref{swfork}), and Eq.(\ref{kondof}) for $T_K$ reduces to
Eq.(\ref{T1}).
 In the limit $\bar d \ll \bar d_{cr}$,
the ground state of the TQD is the non-bonding state $E_{Dn}$ with
an empty site 3 in accordance with the form of the wave function
$|Dn\sigma\rangle$ (\ref{func2}). The SW Hamiltonian
(\ref{swfork}) contains in this case only the term $\sim J_{nnoo}$
and $T_K = \bar D \exp (-{1/j_n})$. The Kondo temperature as a
function of $\bar D $ has a maximum in a crossing point $\bar
D_{cr}$, but unlike the case of cross geometry (Fig.
\ref{kondot}), $T_K(\bar D)$ is nonzero on both sides of this
maximum (Fig. \ref{tk-fork}).

The change of the ground state wave function from the bonding
combination $|D_{b\sigma}\rangle$ to the non-bonding one
$|D_{n\sigma}\rangle$, influences the behavior of tunnel
conductance. Let us compare the tunnel current between the leads
'2' and '1' and between the leads '3' and '1', which is defined by
the components $G_{22}$ and $G_{33}$ of the three-terminal
conductance matrix $G_{ij}=\partial I_i/\partial V_{j}$.

When calculating these components as a function of $\bar D$, one
immediately sees that the Kondo-type ZBA in $G_{22}$ exists in all
three regimes, and the peak of this conductance follows the
behavior of $T_K(\bar D)$. The behavior of $G_{33}$ is more
peculiar. The ZBA in this channel exists only until the even
components $|Db\sigma\rangle$ are involved in Kondo tunneling. In
the limit $T_K \lesssim E_{Db}-E_{Dn}$ at $\bar d \ll \bar d_{cr}$
this anomaly disappears, so we encounter a unique situation where
the Kondo resonance is absent in conductance in spite of the
presence of Kondo screening.

However, the resonance Kondo regime in $G_{33}$ arises as a finite
bias anomaly (FBA). To describe this tunneling one has to retain
the terms $\sim J_{bnoe}$ and $J_p$ in the Hamiltonian
(\ref{swfork}). These terms describe {\it inelastic} tunneling,
which acquires a form of Kondo resonance at finite bias in
accordance with the mechanism offered in Ref. \onlinecite{KMK}. In
that case the tunneling through the double quantum dot with even
occupation in a singlet ground state was considered, and the Kondo
regime becomes relevant at finite bias, when the difference in the
chemical potentials of source and drain leads compensates the
exchange gap between the ground state singlet and excited spin
triplet state. In our case the FBA arises at odd occupation when
the bias compensates the energy gap $\delta=E_{Db}-E_{Dn}$.

\begin{figure}[htb]
\centering
\includegraphics[width=65mm,height=35mm,angle=0]{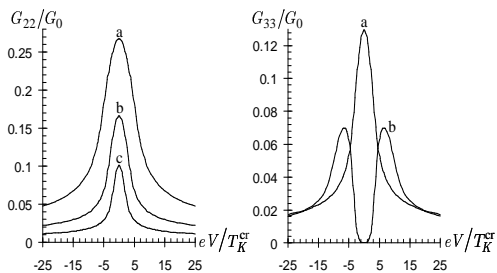}
\caption{Left panel: Tunnel conductance in the channel '2--1' for
$\bar D=\bar D_{cr}$ (a), $\bar D>\bar D_{cr}$ (b) and $\bar
D<\bar D_{cr}$ (c). Right panel: Tunnel conductance in the channel
'3--1' for $\bar D>\bar D_{cr}$ (a) and $\bar D<\bar D_{cr}$ (b).}
\label{cond-fork}
\end{figure}

In order to calculate the tunnel conductance in the weak coupling
regime, we use the modified Golden Rule formula\cite{KNG}, which
reads for the channel '3--1' as $G_{33}(eV_3, T)/G_0\sim
|\widetilde J_{13}(eV_3, T)|^{2}$, where $G_0=e^2/\pi\hbar,$ and
$\widetilde J_{13}$ is the solution of the RG flow equations
(\ref{scal-f}) for $D={\rm max}\{eV_3,T\}$. In case of Kondo-type
resonance at finite bias $eV_3$,  the equation for conductance in
the vicinity of FBA reads \cite{KMK}
\begin{equation}
G/G_0 \sim \ln^{-2}({\rm max}\{eV_3-\delta,T\}/T_K).
\end{equation}
Similar equation (with $V_2\to 0$) describes the ZBA in tunnel
conductance in the channel '2--1'.

The results of calculation of $G_{22}$ and $G_{33}$ are presented
in Fig. \ref{cond-fork}. One may see from this figure how the ZBA
in $G_{22}$ goes through the maximum at the crossing point $\bar
D_{cr}$ whereas the ZBA in $G_{33}$ transforms into FBA for small
enough $\bar D$ when $\delta$ exceeds $T_K$.

Discussion of damping effects, which, to a certain degree, tend to
smear the FBA \cite{KMK,Paas} is beyond the scope of this work. It
should be mentioned, however, that it was shown in Ref.
\onlinecite{KMK} that there exists a wide enough window of
parameters, where this damping is not fatal for existence of well
shaped FBA. It is worth noting also that the tunnel conductance
between the leads '1' and '2' exists in spite of the absence of
direct tunneling channel. The tunneling mechanism is connected in
this case with the "pendulum" structure of the electron wave
function (\ref{func2}) in TQD. In case of ground state $E_{Db}$
this is the bonding combination
$(d^{\dag}_{1\sigma}+d^{\dag}_{2\sigma})|0\rangle$, in case of
ground state $E_{Dn}$ this is the resonating valence bond
$(d^{\dag}_{1\sigma}-d^{\dag}_{2\sigma})|0\rangle$. In the latter
case the dot '3' is excluded from cotunneling.  It is therefore
involved only in the determination of the Kondo temperature.

\section{Concluding remarks}

We have shown in this paper that triple quantum dots in some
special geometries demonstrate unusual behavior in the Kondo
tunneling regime. This behavior stems from  inequivalence of
constituents (side valleys and central valley). Asymmetric double
quantum dot considered in Ref. \onlinecite{KAv} was the first
example of complex quantum dot with inner and outer "shell".
Trimers with inequivalent side and central valleys studied in
Refs. \onlinecite{KuKA,Marcus,Vava} and in the present paper are
more complicated  examples of artificial molecules with shell
structure. In this case the two side dots form an "inner shell"
whereas the big central dot plays the role of an "outer shell". If
the outer shell is open\cite{Marcus,Vava}, it contributes to the
indirect exchange between the electrons in the inner shell.

We considered here the case of closed outer shell and found that
such trimer with odd electron occupation $N=1,3$ possesses
properties, which were observed earlier in dots with even
occupation $N=2$. In particular, the Kondo tunneling may be absent
in the ground spin doublet state of TQD due to special symmetry
properties of the wave function (odd $l-r$ symmetry in case of
$N=3$ and empty outer shell in case of $N=1$). Involvement of the
low-lying Kondo-active spin doublet  results in a two-stage Kondo
screening reminiscent of that found in quantum dot with occupation
$N=2$ where the spin excitation spectrum is formed by the
singlet-triplet pair.\cite{Hof04} Other interesting possibilities
now open due to the resonance valence bond structure of the
electron wave function (\ref{func2}) in case of partially occupied
inner shell in a fork geometry with $N=1$. In particular the
"pendulum effect"\cite{Sarag,Lehur} perceived in TQD with even
occupation $N=2,4$ may be exploited in this type of TQD as well.

The transformation of ZBA into FBA under changing gate voltage is
 a special manifestation of general phenomenon, known as
 "critical phase transition", where
the symmetry of the ground state changes as a function of a
control parameter. Similar effect should be observed in planar and
double quantum dots \cite{KMK}  with even occupation where the
singlet-triplet crossover may occur with changing gate voltage or
in transition metal molecular complexes \cite{KKW}. In the latter
case local phonons are essentially involved in this transition.
\cite{KKW2}

\noindent This work is partially supported by grant from the
Israel Science Foundation (ISF) and the Deutsche Israel Project
(DIP).

\end{document}